\documentclass[english,aps,twocolumn,groupedaddress]{revtex4}
\usepackage[T1]{fontenc}
\usepackage[latin1]{inputenc}
\usepackage{array}
\usepackage{amsmath}
\usepackage{graphicx}

\makeatletter

\providecommand{\tabularnewline}{\\}

\usepackage{array}

\makeatletter

\usepackage{epsf}

\makeatother

\usepackage{babel}
\makeatother
\begin{document}

\title{Experimental Observation of the Inverse Proximity Effect in Superconductor/Ferromagnet
Layered Structures}

\author{R. I. Salikhov, I. A. Garifullin}

\email{ilgizgarifullin@yahoo.com }

\author{N. N. Garif'yanov}

\affiliation{Zavoisky Physical-Technical Institute, Kazan Scientific Center of
Russian Academy of Sciences, 420029 Kazan, Russia}

\author{L. R. Tagirov}

\affiliation{Kazan State University, 420008 Kazan, Russia}

\author{K. Theis-Bröhl, K. Westerholt, H. Zabel}

\affiliation{Institut für Experimentalphysik/Festkörperphysik, Ruhr-Universität
Bochum, D-44780 Bochum, Germany }

\date{\today}

\begin{abstract}
We have studied the nuclear magnetic resonance (NMR) of $^{51}$V
nuclei in the superconductor/ferromagnet thin film heterostructures
Ni/V/Ni and Pd$_{1-x}$Fe$_{x}$/V/Pd$_{1-x}$Fe$_{x}$ in the normal
and superconducting state. Whereas the position and shape of the NMR
line in the normal state for the trilayers is identical to that observed
in a single V-layer, in the superconducting state the line shape definitely
changes, developing a systematic distortion of the high-field wing
of the resonance line. We consider this as the first experimental
evidence for the penetration of ferromagnetism into the superconducting
layer, a phenomenon which has been theoretically predicted recently
and dubbed the \textit{{inverse proximity effect. }}
\end{abstract}
\maketitle The penetration of the superconducting condensate into
a ferromagnet in superconductor/ferromagnet (S/F)
heterostructures, the so-called S/F proximity effect, is well
understood by now (see, \textit{e.g.,} \cite{Buzdin} and
\cite{Efetov}) for recent reviews. However, only recently it has
been shown theoretically that the inverse effect, namely the
penetration of ferromagnetism into a superconductor, should also
exist \cite{Bergeret,Kharitonov}. The basic physical origin for
this so-called \textit{inverse proximity effect} can qualitatively
be understood rather easily: The conduction electron spins of a
thin F-layer are predominantly oriented in one direction because
of the conduction electron spin polarization. Due to
superconducting correlations at distances of the order of the
Cooper pair size $\xi_{s}$ from the S/F interface, these electrons
have Cooper partners with the opposite spin direction in the
S-layer, thus giving rise to a spin polarization of the conduction
electrons in the S-layer with a direction opposite to the
ferromagnet.

Until now there is no unequivocal experimental evidence for the
inverse proximity effect in the literature. A quantitative
estimation shows that the amplitude of the induced magnetization
in the S-layer is small, especially when taking the limited
transparency of the S/F interface in a real thin film system into
account \cite{Efetov}. Thus one needs an experimental method with
very high sensitivity for small changes of the spin polarization
in the S-layer. The induced spin polarization in the
superconductor corresponds to a decrease in the spin
susceptibility of the conduction electrons upon the transition to
the superconducting state. The spin susceptibility of the
conduction electrons, on the other hand, is one of the physical
reasons for the Knight shift of the nuclear magnetic resonance
(NMR) line in metals. Thus, in NMR the inverse proximity effect
should manifest itself as a decrease of the Knight shift upon the
transition to the superconducting state. As will be shown below,
using the super sensitive NMR techniques, it actually is possible
to observe this effect.

The choice of a suitable F/S material combination is of primary importance
for an observation of the inverse proximity effect. The material for
the S-layer should have a strong NMR signal, a high superconducting
transition temperature and a high quality interface with the F-material.
In addition, there should be an appreciable change of the Knight shift
on the transition to the superconducting state. Among the suitable
elemental superconductors Pb, Nb and V appear as the most promising
candidates \cite{Schrieffer,MacLaughlin}. However, only V fulfills
the condition of a high interface quality with epitaxial growth on
Fe \cite{Hjorvarsson} and high interface transparency for the electrons
\cite{Garifullin}. Recently we have shown that the Knight shift of
pure V changes upon the transition to the superconducting state \cite{JETPLett},
very similar to pure Nb \cite{Rossier}, which has a similar electronic
structure. This is contrary to earlier experiments of Noer and Knight
\cite{Noer}.

In order to reach a noticeable magnetic polarization due to the inverse
proximity effect, the S-layer thickness in the S/F heterostructures
should match the superconducting coherence length. In our case it
is restricted to about 30 nm (see Table 1), implying that the number
of V-nuclei involved in the resonance signal will be extremely small.
Conventional NMR techniques then encounter serious sensitivity problems.
In order to overcome this, we built a super sensitive NMR spectrometer
operating in a continuous mode at the frequency of about 6 MHz \cite{1},
based on a Robinson-scheme generator (see, \textit{e.g.}, \cite{Wilson}).
MOSFET transistors capable of operating at 4 K enable the immersion
of the HF generator into liquid helium in close vicinity to the sample
pick-up coil. This strongly reduces thermal noise and excludes losses
in the connecting line. Since the gyromagnetic ratio for the Cu and
V nuclei are very similar, the resonator coil as well as the magnetic
field modulation coils were wound of high-purity silver wire. At liquid-helium
temperatures, the resonance circuit has a high $Q$-value that also
considerably enhances the NMR spectrometer sensitivity. The generator
output was connected to a lock-in amplifier working with digital signal
processing. For the generation of the dc magnetic field we used the
magnetic system of the Bruker EPR spectrometer ER-418, which includes
field sweep option and stabilization by a Hall unit. Precision measurements
of the magnetic field were performed by a NMR gaussmeter whose NMR
sensor was always in a strictly fixed position. The experimental error
in the measurement of the magnetic field including its inhomogeneity
in the operating range ($4\times10^{-5}$ Oe/cm) did not exceed 0.2
Oe.

For the NMR study we have prepared F/S/F trilayers using V as the
superconducting layer and either Ni or an alloy Pd$_{1-x}$Fe$_{x}$
as the ferromagnetic layers (see Table 1). All layers were grown on
single-crystalline MgO(001) substrates by molecular beam epitaxy in
a growth chamber with a base pressure below $5\times10^{-10}$ mbar.
For V, Ni and Pd we used electron beam evaporation and a growth rate
of 0.15, 0.03 and 0.05 nm/s, respectively. The Pd$_{1-x}$Fe$_{x}$
alloys were produced by co-evaporation of Pd and Fe. Fe was evaporated
from an effusion cell with an evaporation rate depending on the desired
concentration of Fe in the Pd$_{1-x}$Fe$_{x}$ alloy. The concentrations
of Fe in the Pd$_{1-x}$Fe$_{x}$ alloys given in Table 1 were refined
with the data for $T_{Curie}=f(x)$ taken from literature \cite{PdFeCurie}.
From the temperature dependence of the magnetization measured using
a SQUID magnetometer we derived a ferromagnetic Curie temperature
$T_{Curie}\simeq$75 and 100 K for the two samples of Table 1, for
these Curie temperatures we estimate an Fe concentration $x$ of 0.02
and 0.03 (see Table 1). To prevent oxidation, all samples were capped
by Pd layers. A growth temperature of 300$^{o}$ C was chosen for
all layers, this temperature provides a good compromise between crystallinity
and low interdiffusion at the interfaces. \textit{In-situ} reflection
high energy electron diffraction during the growth process revealed
smooth layer growth of all layers. The thickness and the quality of
the films were investigated by conventional small-angle x-ray reflectivity.
Well resolved Kiessig fringes from the total layer thickness were
clearly observed. Fits using the modified Parratt formalism \cite{Parratt,Nevot}
gave the thickness of the V-layer $d_{V}$ and the interface roughness
parameter (Rough), also included in Table 1. Due to the fact that
the structural quality of V grown directly on MgO is much higher than
for the growth on Ni and Pd$_{1-x}$Fe$_{x}$ layers, the Rough values
for the single V layer and the trilayer systems are different (see
Table 1, column 3).

The upper critical field $H_{c2}^{\perp}(T)$ for the magnetic
field direction perpendicular to the film plane (Fig.1) has been
measured resistively by the standard four point dc technique. The
superconducting transition temperature $T_{c}$ of the samples from
Table 1 is between 3.6 K and 5 K (see fourth column of Table 1).
From the residual resistivity ratio $RRR=R(300K)/R(5K)$ (fifth
column Table 1) we can determine the residual resistivity
$\rho_{0}$ by using the phonon contribution to the resistivity of
V, $\rho_{phon}$(300 K) = 18.2 $\mu\Omega\cdot cm$. Following
Lazar \textit{et al.} \cite{Lazar} and using the Pippard relations
\cite{Pippard}, we get $\rho_{0}l=2.5\times10^{-6}\ \mu\Omega\cdot
cm$$^{2}$ and can calculate the mean free path $l$ of the
conduction electrons (6th column, Table 1). The BCS coherence
length for V is $\xi_{0}$ = 44 nm. A comparison of $l$ and
$\xi_{0}$ implies that the superconducting properties of our
samples are closer to the \char`\"{}dirty\char`\"{} limit
($l<<\xi_{0}$) than to the \char`\"{}clean\char`\"{} limit
($l>>\xi_{0}$). In the \char`\"{}dirty\char`\"{} limit
$\xi_{s}=\sqrt{\xi_{0}l/3.4}$ holds, which is given in the last
column of Table 1.

The NMR measurements were performed on the $^{51}$V nuclei in the
temperature range from 1.4-4.2 K. Since the operating frequencies
are slightly different for different samples, in order to compare
the resonance line positions directly, all data were recalculated
to the same frequency $\nu$, in our case to $\nu$ = 5542.3 kHz.
All measurements were performed with a perpendicular orientation of
dc magnetic field relative to the film plane. The signal-to-noise
ratio does not exceed 3 and therefore we accumulated signals from
at least 20$\div$30 sweeps of the magnetic field during two minutes
each.

In the normal state, the $^{\text{51}}$V NMR line shape and
position were approximately the same for all samples studied.
Therefore, in Fig. 2a we show the NMR signal in the normal state
for the single V-layer only. The resonance line shape is described
by the derivative of a Gaussian absorption curve. Fitting this
theoretical curve we can determine the resonance line position
with an absolute accuracy below 0.5 Oe. For the resonance line
width (the peak-to-peak distance of the absorption line
derivative) we get a value of 12.2 Oe. The resonance field at
$H_{0}$ = 4923.1 Oe is found to be shifted by $\delta H$ = 29.1 Oe
relative to the position in an insulator (4952.2 Oe for $^{51}$V),
thus, for the Knight shift in the normal state, which is defined
as the ratio of the NMR line shift relative to its position in an
insulator, we get 0.59 $\pm$0.01 \%, in good agreement with the
value measured previously \cite{Noer,JETPLett}.

When decreasing the temperature below the superconducting transition
temperature for the single V-layer (Fig. 2b), the resonance line shifts
to higher magnetic fields with some broadening of the line width,
but the resonance line shape can still be described by a Gaussian
absorption curve. For the F/S/F trilayer samples (Fig. 2c-2e), however,
after the transition to the superconducting state, the NMR line shape
is found to be markedly changed, the high-field wing of the NMR line
appearing strongly distorted. As will be discussed next, this can
be taken as an experimental manifestation of the inverse proximity
effect.

In NMR experiments in an external magnetic field $H_{e}$
perpendicular to the film plane the superconducting film of a type
II superconductor is in the vortex state, and the resonance line
is shifted and broadened due to the intrinsic spatial
inhomogeneity of the magnetization in the vortex state. For a
detailed theory of NMR in type II superconductors in the vortex
state see \cite{Rossier,Delriew}. Following these calculations,
the magnetic field in the center of the vortex core is given by
$H_{v}=H_{e}-4\pi M$, ($M$ is the magnetization). At the saddle
point of the vortices for a triangular vortex lattice in the dirty
limit the result for the local magnetic field is
$H_{S}=H_{e}+0.34\times4\pi M$ \cite{Dobrosavljevic}. The portion
of nuclei in the Abrikosov vortex lattice is maximum at the saddle
point $H_{s}$ (see, e. g. \cite{Rossier}). Thus if the
magnetization is not too large (i.e. if the resonance field is
close to the upper critical magnetic field $H_{c2}$) one expects
that the NMR resonance field shifts a bit to the high-field side
and is slightly broadened, but still can be approximated by a
Gaussian absorption curve. For the pure V-film (Fig. 2a,b) this is
just what we observe, the shift of the resonance field being about
9 Oe, similar to what we have found \cite{JETPLett} in bulk V
before. However, for the F/S/F trilayers in Fig. 2 c-e the line
shape is definitely changed in the superconducting state. It is
asymmetric with a strongly distorted high-field wing, suggesting
that in the superconducting state there is an additional mechanism
determining the line shape.

As discussed above, according to the model of the inverse
proximity effect \cite{Bergeret}, spin-polarized electrons from
the interfacial region penetrate into the superconducting layer.
By means of the hyperfine interactions this spin-polarization
induces a local field $H_{loc}$ on the V nuclei with a direction
opposite to the external magnetic field (we suppose that the
conduction electron spin polarization in the ferromagnetic layer
is in the direction of the applied field) and the NMR resonance
field shifts to higher fields accordingly. In quantitative terms,
in order to calculate the NMR line shape, one must also take the
spatial distribution of the spin polarization in the
superconducting layer into consideration. The induced
spin-polarization $\sigma(x)$ in the superconductor is expected to
decay exponentially with the distance \textit{x} from the
interface \cite{Bergeret}. Then, the corresponding local field on
a nucleus is \begin{equation}
H_{loc}=H_{m}exp(-x/\xi_{s})\propto\sigma(x),\end{equation}
 where $x$=0 corresponds to the position of the F/S interface \cite{Bergeret}.
The local field distribution, \begin{equation}
F(H)={\frac{1}{{d}}}\int_{0}^{d}dx(H-H_{loc}(x)),\end{equation}
 has to be convoluted with the unperturbed NMR Gaussian line shape
derived by fitting of the normal-state NMR line.

The result of a numerical simulation of the NMR line shape in a
superconducting film with a finite spin polarization penetrating
from the edges is shown in Fig.~3. The line clearly exhibits a
broadened high-field wing, strikingly similar to the experimental
spectra observed for PdFe/V/PdFe and Ni/V/Ni trilayers in Fig. 2.
The low-field side of the resonance line is mainly determined by
the V nuclei in the middle of the V-layer, and the shape remains
essentially unaffected by the spin polarization. The high-field
side, however, is modified, since here the V nuclei from the
region close to the S/F interfaces sensing a local field from the
spin polarization contribute to the NMR signal. The degree of
distortion of the high-field wing of the NMR resonance line should
scale with the amplitude of the spin polarization This trend is
clearly present in the sequence of the experimental spectra from
Fig. 2c-e. We did not superimpose the calculated lineshape on the
experimental curves, because we do not have a realistic
quantitative estimate for the conduction electron spin
polarization (and consequently the parameter $H_{m}$ in formula
(1)).

In summary, the character of the NMR line distortion and the systematic
increase of the distortion with the strength of the ferromagnet below
the superconducting transition temperature in the F/S/F trilayers
leads us to the conclusion that we have observed a manifestation of
\textit{ferromagnetism penetrating into the superconductor,} i.e.
the novel mechanism coined the \textit{inverse proximity effect} in
Refs. \cite{Bergeret} and \cite{Kharitonov}.

{\small We are grateful to Professor Konstantin B. Efetov and
Anatoly F. Volkov for stimulating this study. This work was
supported by the Deutsche Forschungsgemeinschaft within the SFB
491 and by the Russian Foundation for Basic Research (project nos.
08-02-00098 (experiment) and  07-02-00963(theory)).}{\small \par}


\bigskip{}

TABLE 1. Experimental parameters of the studied samples. Given are
the thickness of the V layer $d_{V}$, the roughness parameter
Rough obtained from the fit of the small angle x-ray reflectivity,
the superconducting transition temperature $T_{c}$, the residual
resistivity ratio $RRR$, the electron mean free path in the V
layer $l$, and the superconducting coherence length $\xi_{s}$. The
thickness of the magnetic layers is about 3 nm for all trilayer
samples.

\vspace{15mm}

\begin{tabular}{|p{2.5cm}|c|c|c|c|c|c|c|}
\hline Sample & $d_{V}$ & Rough & $T_{c}$ & $RRR$ & $l$ &
$\xi_{s}$& \tabularnewline \hline & (nm) & (nm) & (K) & & (nm) &
(nm)& \tabularnewline \hline V & 30 & 0.3 & 4.65 & 11 & 15 & 14 &
\tabularnewline \hline Pd$_{0.98}$Fe$_{0.02}$/V/ \newline
Pd$_{0.98}$Fe$_{0.02}$ & 36 & 1.3 & 3.02 & 4.6 & 5 & 8 &
\tabularnewline \hline Pd$_{0.97}$Fe$_{0.03}$/V/ \newline
Pd$_{0.97}$Fe$_{0.03}$ & 42 & 1.3 & 3.55 & 6 & 7 & 10 &
\tabularnewline \hline Ni/V/Ni & 44 & 1.6 & 4.05 & 4.4 & 5 & 8 &
\tabularnewline \hline
\end{tabular}\\

\bigskip{}

\begin{figure}[p]
 \centering{\includegraphics[width=7cm]{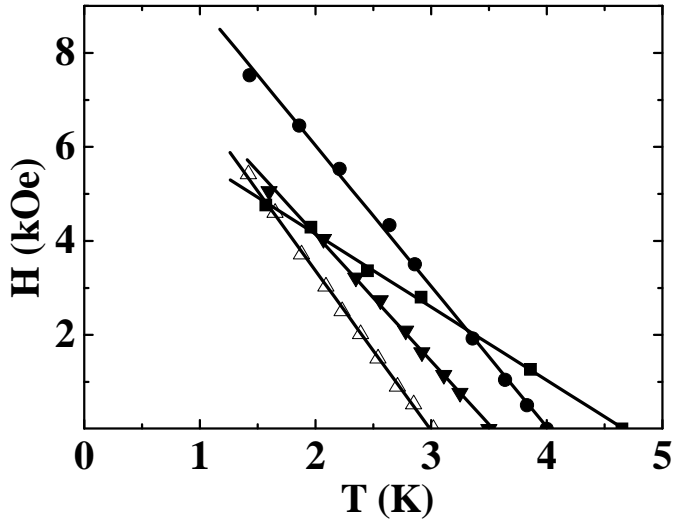}}

\caption{The upper critical field vs temperature for studied
samples in the perpendicular orientation of the external magnetic
field relative to the film plane. Full squares relate to the
single V-layer, open triangles - to
Pd$_{0.98}$Fe$_{0.02}$/V/Pd$_{0.98}$Fe$_{0.02}$, closed
triangulares - to Pd$_{0.97}$Fe$_{0.03}$/V/Pd$_{0.97}$Fe$_{0.03}$,
and closed circles - to Ni/V/Ni trilayers.}
\end{figure}

\begin{figure}
\centering{\includegraphics[width=7cm]{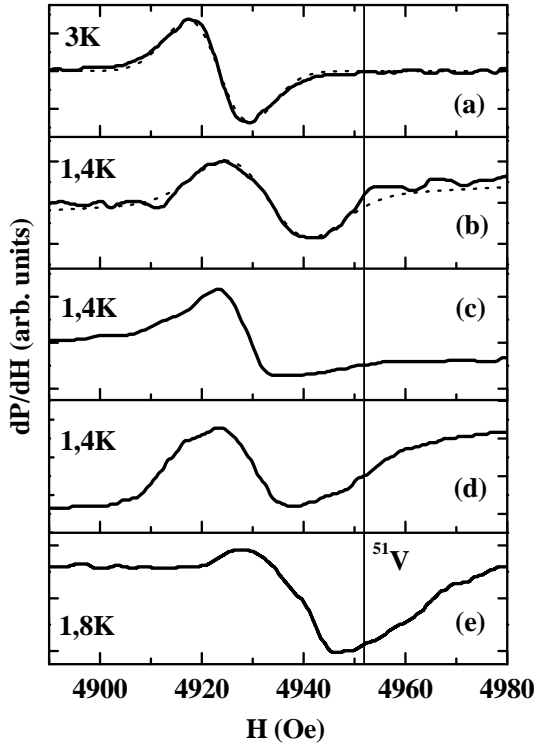}}

\caption{NMR spectra for the single V-layer in the normal ($T$ = 3
K) (a) and superconducting ($T$ =1.4 K) (b) states and for
Pd$_{0.98}$Fe$_{0.02}$/V/Pd$_{0.98}$Fe$_{0.02}$ (c),
Pd$_{0.97}$Fe$_{0.03}$/V/Pd$_{0.97}$Fe$_{0.03}$ (d) and Ni/Pd/Ni
(e) trilayers in the superconducting state. All data are given for
the external magnetic field perpendicular to the sample plane. The
NMR spectra for the single V layer are simulated with the Gaussian
lineshape of equivalent peak-to-peak linewidth (dashed curves).}
\end{figure}

\begin{figure}[ht]
 \centering{\includegraphics[width=7cm]{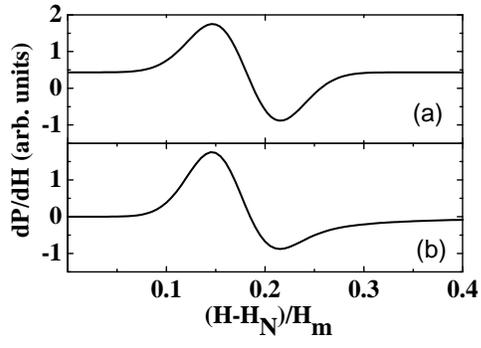}}

\caption{(a) The pure Gaussian lineshape derivative with
peak-to-peak width and position equivalent to the model spectrum
below. (b) Model calculations of the NMR line shape in an F/S/F
trilayer with $\xi_{s}/d$=0.2 and the Gaussian broadening
parameter $\sigma/H_{m}$=0.06 (b). $H_{N}$ is the resonance field
in the normal state. Only the lineshape distortion and the line
shift due to the inverse proximity effect were considered in
calculation of the spectrum (b).}
\end{figure}


\begin{thebibliography}{10}
\bibitem{1} The resonace field should be smaller than the critical
field of superconductor.
\end{thebibliography}
\end{document}